\documentclass[aps,pra,twocolumn,showpacs,superscriptaddress,amsmath,amssymb]{revtex4}
\usepackage{graphicx,epsfig}
\begin{document}


\title{Relativistic Tunneling Through Two Successive Barriers}



\author{Jos\'{e} T. Lunardi}
\email[corresponding author: ]{jttlunardi@uepg.br}
\thanks{\emph{Presently at Dipartimento di Fisica e Tecnologie
Relative, Universit\`a di Palermo. Viale delle Scienze, Edificio 18.
I-90128 Palermo, Italy}. }
\affiliation{Grupo de F\'{\i}sica Te\'orica e Modelagem
Matem\'atica, Departamento de Matem\'atica e Estat\'{\i}stica,
Universidade Estadual de Ponta Grossa. Av. General Carlos
Cavalcanti, 4748. CEP 84032-900, Ponta Grossa, PR, Brazil}

\author{Luiz A. Manzoni}
\email[]{manzoni@macalester.edu}
\affiliation{Department of Physics and Astronomy, Macalester
College, 1600 Grand Avenue, Saint Paul, MN 55105, USA}


\date{\today}

\begin{abstract}
We study the relativistic quantum mechanical problem of a Dirac
particle tunneling through two successive electrostatic barriers.
Our aim is to study the emergence of the so-called \emph{Generalized
Hartman Effect}, an effect observed in the context of
nonrelativistic tunneling as well as in its electromagnetic
counterparts, and which is often associated with the possibility of
superluminal velocities in the tunneling process. We discuss the
behavior of both the phase (or group) tunneling time and the dwell
time, and show that in the limit of opaque barriers the relativistic
theory also allows the emergence of the Generalized Hartman Effect.
We compare our results with the nonrelativistic ones and
discuss their interpretation.

\end{abstract}


\maketitle


\section{Introduction}

The phenomenon of tunneling is one of the most striking and extensively
studied consequences of quantum mechanics. Yet, after decades of scrutiny
(for reviews see, for example, \cite{HSt89, LMa94, Win06-2, Win06-1}) it still
presents serious conceptual challenges, such as a meaningful definition
of tunneling time, that is, the
time it takes for a particle to tunnel through a potential barrier.

Several different scales of time associated with the tunneling
process have been proposed (see \cite{HSt89}). Among the most
prominent ones are the phase time (or group delay time), given by
the energy derivative of the phase shift in the transmission (or
reflection) amplitude, and the dwell time, which is related to the
average time spent by the particle in the region of the potential.

It is well known that for the tunneling of a particle through an
opaque barrier the group delay saturates with the width of the
barrier, a phenomenon  called {\it Hartmann effect} \cite{Har62}.
Several authors interpret this saturated time as the transit time
for the particle to go through the potential, which would imply, as an
immediate consequence, the possibility of superluminal (group)
velocities for barriers with a sufficiently large spatial extension.
Such an interpretation has been in the center of an intense debate
in the literature (see, e.g., \cite{Win06-1, Win03-1, NHa02} and references there cited).

Recently an apparently even more paradoxical effect, which became
known as {\it Generalized Hartman Effect}, has been brought to
attention, not only in the context of nonrelativistic quantum
tunneling, but also in the context of its electromagnetic
counterparts. This effect consists in the fact that for tunneling
through two potential barriers separated by a distance $l$ the phase
time is, in the limit of opaque barriers, independent not only on
the barrier widths but also on the spacing {\it between} them
\cite{ORS02} (see also \cite{LLB02, Win05}). In fact, Esposito \cite{Esp03}
showed that for a system of $N$ barriers the phase time is
independent also on the number of barriers. Despite the fact that
phase time cannot, in general, be interpreted as a propagation (or
transit) time for the particle (or wavepacket), this effect is
counterintuitive since one would, \textit{naively}, expect that in
the space between the barriers the group delay could be viewed as a
propagation time, and therefore, it should depend on the distance
between the barriers.

In the last years several papers have also analyzed the problem of
quantum tunneling from a relativistic
standpoint \cite{LSM98, LCh02, ZLo07, KSG01, PJa03, CLi03, WNL04, LRo07}. Some of these papers
were concerned with the fact that the analysis of possible
\textit{superluminal} group velocities associated with the Hartmann
effect should be properly addressed in the context of a relativistic
theory \cite{LSM98, KSG01, PJa03, CLi03, LRo07};
others were concerned with general aspects of the relativistic
problem, such as the relation between phase time and dwell
time \cite{WNL04} or the relation between dwell time and Larmor
times \cite{ZLo07}. In fact, due to the relevance of the tunneling
phenomenon, it is important to consider the possible quantitative
and/or qualitative differences in the phase (and dwell) time arising
due to the relativistic dynamic. What is more, a clear
understanding of the relativistic aspects of tunneling is imperative
if one wants to eventually obtain a meaningful time scale for this
phenomenon, because the instantaneous spread of the probability
density in nonrelativistic quantum mechanics \cite{Got} makes it
difficult to define an unambiguous tunneling time in the context of
Schr\"{o}dinger theory.

Most of the above papers were concerned with a single potential
barrier or potential well. In fact, to the best of our knowledge,
the only relativistic treatment of the two barrier case was due to
Leavens  and Aers \cite{LAe89}, which were concerned with Larmor
times at resonance. Thus, the present work complements the previous
ones by considering the relativistic approach to the problem of a
wave packet tunneling through two successive barriers. We shall
address the emergence of the generalized Hartman effect in this
context and discuss its possible interpretations. We also compare
our results with those obtained in the nonrelativistic framework.


\section{The relativistic phase and dwell times}

Here we shall be concerned with the relativistic one-dimensional
scattering of a mass $m$ and spin-$1/2$ wave packet by an
electrostatic time-independent potential $V(z)$. The general form of
the incident wave packet is given by
\begin{equation}
\Psi (z, t) = \int dE \;A(E) \psi (z) e^{-i\frac{Et}{\hbar}}\, ,
\label{wavepack}
\end{equation}
where $A(E)$ are its Fourier coefficients, while $\psi (z)$ must
satisfy the time-independent Dirac equation associated with the energy
$E$,
\begin{equation}
\left\{ -i\hbar c\alpha^z \partial_z +\beta mc^2 + V(z) \right\}
\psi (z) = E \psi (z)\, , \label{dirac}
\end{equation}
with $\alpha^z$ and $\beta$ being the Dirac matrices (we will follow
the conventions in \cite{BDr64}).
The potential in which we are interested consists of two potential
barriers of height $V_0$ and width $a$, and spaced by a distance
$l$, as seen in Figure \ref{fig1} (since barriers of different
heights and widths do not introduce any novelty, they will not be
considered here).

Moreover, we shall consider an incident wave packet whose Fourier
energy distribution $A(E)$ is very sharply concentrated around a
given value $E_0$, corresponding, therefore, to a smooth modulation
of the eigenfunction corresponding to $E_0$. Such a wave packet must
have, therefore, a large spatial extension. For our purposes we
shall assume that the energy dispersion of the wavepacket is
sufficiently narrow such that its spatial extension is always very
large when compared to the extension of the region in which the
potential is nonvanishing (the region $0<z<l+2a$ in Figure
\ref{fig1}). With these assumptions in mind it is justifiable the
use of the stationary phase method to follow the position of the
peak of the wavepacket in the free regions (regions $I$ and $V$)
(see, for example, \cite{statphase} and references therein). We
shall also consider that $E_0$ is a positive energy (particle) in
the evanescent region, $E_0-mc^2<V_0<E_0+mc^2$. Therefore, the
region of supercritical potential, in which there is pair production
(and the associated Klein paradox) and where, therefore, the
one-particle Dirac equation ceases to be valid, will not be
considered (for a study of the supercritical region for the one
barrier potential see \cite{LRo06}).

Considering, as usual, a wave packet incident only from the left and
having spin up (we make this assumption without loss of generality,
because the potential considered here causes no spin flip), the
general solution of the stationary problem in the various regions
indicated in Figure \ref{fig1} is given by
\begin{eqnarray}
&&\psi_I (z) = e^{ikz} u_E(k) + R e^{-ikz} u_E(-k)\, , \nonumber \\ \nonumber \\
&&\psi_{II} (z) = A e^{-qz} u_{E-V_o}(iq) + B e^{qz} u_{E-V_o}(-iq) \, ,\nonumber \\ \nonumber \\
&&\psi_{III} (z) = C e^{ikz} u_E(k) + D e^{-ikz} u_E(-k) \, , \\ \nonumber \\
&&\psi_{IV} (z) = F e^{-qz} u_{E-V_o}(iq) + G e^{qz} u_{E-V_o}(-iq) \, ,\nonumber \\ \nonumber \\
&&\psi_V (z) = T e^{ikz} u_E(k)\, ,\nonumber
\end{eqnarray}
where
\begin{equation}
u_E(k)= \left(
              \begin{array}{c}
                       1 \\
                       0 \\
                       \frac{ck\hbar}{E+mc^{2}} \\
                       0 \\
                     \end{array}
                   \right)
\end{equation}
and
\begin{equation}
k\equiv\frac{1}{\hbar c}\sqrt{E^{2}-m^{2}c^{4}}\; ; \hspace{0.5cm}
q\equiv\frac{1}{\hbar c}\sqrt{m^{2}c^{4} - (E-V_{0})^{2}}\, .
\end{equation}
The above coefficients can be determined, as usual, from the
boundary conditions requiring the wave function to be continuous at
the potential discontinuities. After some simple but tedious algebra
we obtain the transmission and the reflection amplitudes:
\begin{figure}
\includegraphics[width=7cm,height=4.0cm]{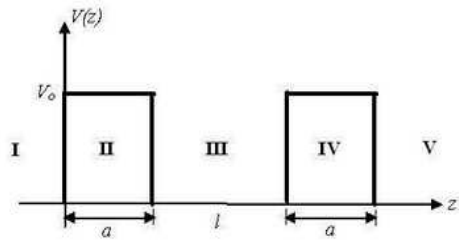}
\caption{Schematic view of the two barriers arrangement. We consider
two rectangular identical barriers of heigth $V_0$ and width $a$,
separated between themselves by a distance $L$.\label{fig1}}
\end{figure}
\begin{eqnarray}
T&=&  e^{-2ika}\left\{ \left[\cosh(qa) +
i\frac{(1-\alpha^{2})}{2\alpha}\sinh(qa)\right]^2 \right. \nonumber \\
\nonumber \\
&+&\left.
\frac{(1+\alpha^{2})^{2}}{4\alpha^{2}}\sinh^2(qa)e^{2ikl}\right\}^{-1}
\label{t}
\end{eqnarray}
and
\begin{eqnarray}
R&=& {\mathrm{e}}^{i [k (2a+l)-\frac{\pi}{2}]}\,
\frac{(1+\alpha^2)}{\alpha} \sinh
(q a)\nonumber \\
&\times& \left\{ \cos (k l)\cosh (q a) +
\frac{1}{2\alpha}(1-\alpha^2)\sin (k l)\sinh(q a)
\right\} \nonumber \\
&\times& T\, , \label{r}
\end{eqnarray}
where we have introduced
\begin{equation}
\alpha\equiv\frac{k}{q}\frac{(E-V_0+mc^2)}{(E+mc^2)}\, .
\end{equation}
It is convenient to express the transmission and reflection
coefficients in terms of their phases as
\newpage
\begin{eqnarray}
T(E) &=& |T|{\mathrm{e}}^{i[\varphi_t-k(2a+l)]}\, ;\label{tt} \\
R(E) &=& |R|{\mathrm{e}}^{i\varphi_r}\, ,\label{tr}
\end{eqnarray}
where $\varphi_r\!=\!\varphi_t\!-\!\pi/2$, while $\varphi_t$ is
given by
\begin{widetext}
\begin{equation}
\varphi_t = kl -\tan^{-1}\left\{
\frac{4\alpha(1-\alpha^2)\sinh(2qa)-(1+\alpha^2)^2\sin(2kl)[1-\cosh(2qa)]}
{4\alpha^2[1+\cosh(2qa)]+[1-\cosh(2qa)]\left[(1-\alpha^2)^2-(1+\alpha^2)^2\cos(2kl)\right]
} \right\}\, .
\label{fase}
\end{equation}
\end{widetext}
Now, accordingly to the stationary phase method, the (extrapolated)
transmitted and reflected phase times are given, respectively, as
\cite{HSt89}
\begin{eqnarray}
\tau_p^t&=&\left.\hbar\frac{d\varphi_t}{dE}\right|_{E_0}\, , \label{tp} \\
\tau_p^r&=&\left.\hbar\frac{d\varphi_r}{dE}\right|_{E_0}=\left.\hbar\frac{d\varphi_t}{dE}\right|_{E_0}\,
,
\end{eqnarray}
where we used in these expressions the same central energy $E_0$ of
the incident wave packet, what is justifiable by our previous
assumptions about the sharp concentration of the initial wave packet
around this energy (this corresponds to the situation in which there
is essentially no distortion nor reshaping of the transmitted wave,
a condition claimed by several authors as necessary to allow a
physical meaning to the group velocity \cite{distortionless,
Win06-1}). The above expressions imply the equality of the
transmitted and the reflected phase times, as it is always the case
for symmetric potentials \cite{falk}. From now on we will refer to
both these times simply as the \emph{phase time} $\tau_p$. Such a
time corresponds to the (extrapolated) instant in which the
transmitted and reflected wave packet \emph{peak} appear at
$z\!=\!2a\!+\!l$ and $z\!=\!0$, respectively.

Now, by using a general relation obtained by Winful \textit{et al.}
\cite{WNL04, Win03}, we can determine the dwell time $\tau_d$,
which is a measure of the time spent by the particle in the
potential region, without distinction of whether it is finally
reflected or transmitted \cite{Smi60, butt}. Such a relation,
for symmetric potentials, reads
\begin{equation}
\tau_d=\tau_p-\tau_i\, ,
\label{dwell}
\end{equation}
where $\tau_i$ is the self-interference delay, given by
\begin{equation}
\tau_i=-\frac{m}{\hbar k^2}
{\mathrm{Im}}(R)\, .
\label{si}
\end{equation}

The explicit expressions obtained for the phase and dwell time from
the above definitions are not particularly illuminating and are
presented in the Appendix. Here we will discuss their properties.
The limit of one barrier (of width $2a$) is easily obtained by
assuming $l=0$ and it agrees with the results of \cite{KSG01} and
\cite{PJa03}. Also the nonrelativistic limit, obtained by making
$mc^2 \rightarrow \infty$ and $V_0 \ll mc^2$, agrees with the
results obtained by \cite{ORS02}. In fact, we can verify these
limits directly in the expressions for the amplitudes and the
transmission phase above.

\section{Discussion and Concluding Remarks}

Of special interest for us is the limit of opaque barriers,
$qa\!\gg\! 1$, in which the phase and dwell times become
\begin{eqnarray}
\tau_p&=&\frac{2\hbar}{1+\alpha^2}\left(
\frac{d\alpha}{dE}\right)=\frac{2\alpha}{1+\alpha^2}
\frac{(k^2+q^2)}{k^2}\frac{m}{\hbar q^2
}\, , \label{p-op}\\
\tau_d&=&\frac{2\alpha}{(1+\alpha^2)}\frac{m}{\hbar q^2}\,
,\label{d-op}
\end{eqnarray}
where we have used the result that in this limit
$\tau_i\!=\!2\alpha/(1\!+\!\alpha^2) m/(\hbar k^2)$. From the above
expressions it is clear that both the phase and dwell times saturate
in the opaque limit, not depending either on the width of the
barriers or on the distance $l$ between them. This demonstrates that
the \textit{generalized Hartmann effect} also emerges in the context
of relativistic quantum mechanics. As a consequence, if we
\emph{extrapolate} the concept of the \textit{group} velocity into
the potential region, it will be given by
$v_g\!\equiv\!(2a\!+\!l)/\tau_p$. This velocity can be made
arbitrarily large, allowing for superluminal \textit{group}
velocities for sufficiently large barrier widths $a$, as we shall
see below.

Figure \ref{figure2}A shows the typical behavior of the phase and
dwell times in the domain of fully relativistic energies, where we
fixed all the relevant parameters, except the barriers width $a$. We
observe that for small values of the width $a$, both the phase and
dwell times are greater than the free and the light times. As the
barrier become thicker, both these times grow slower (in fact, they
can even decrease, depending on the value of the other parameters,
as it is the case shown in the figure) and they can become smaller
than the free and the light times. What is more interesting, the
phase and dwell times can become smaller than the light time
\emph{even before the saturated regime is obtained}. So, in the
domain of fully relativistic energies the group velocity can be
superluminal even before arriving at the opaque limit. A similar
behavior can be observed also in the one barrier case (obtained by
taking $l\!=\!0$). For comparison, we also showed the behavior of
the phase time calculated from Schr\"odinger equation. We see that
the nonrelativistic theory predicts a phase time of the same order
as the relativistic one, and sharing the same behavior, specially in
what concerns the possibility of emerging superluminal group
velocities before the saturation. Figure \ref{figure2}B shows the
same plots for the energies in the relativistic scales, but now with
a greater difference between the energy of the incident packet and
the height of the barrier. Figures \ref{figure2}A and \ref{figure2}B
show that Dirac theory can predict group velocities which can be
\emph{smaller} (Figure 2A) \emph{or greater} (Figure 2B) than those
predicted by the Schr\"odinger theory, being each of these
situations determined by the specific choice of values for the
parameters, especially for the energies. These results are in
agreement with those observed for the single barrier case in
reference \cite{PJa03}, and explain the origin of the apparent
contradictory claims of Krekora \textit{et al}. \cite{KSG01} and
Leavens and Aers \cite{LAe89}, concerning to whether the
relativistic theory predicts group velocities smaller or greater
than those predicted by the nonrelativistic one. In Figure
\ref{figure2}C we can check the complete agreement of the
predictions from both the relativistic and the nonrelativistic
theories in the scale of low (nonrelativistic) energies.
\begin{figure}
\includegraphics[width=9.2cm,height=13cm]{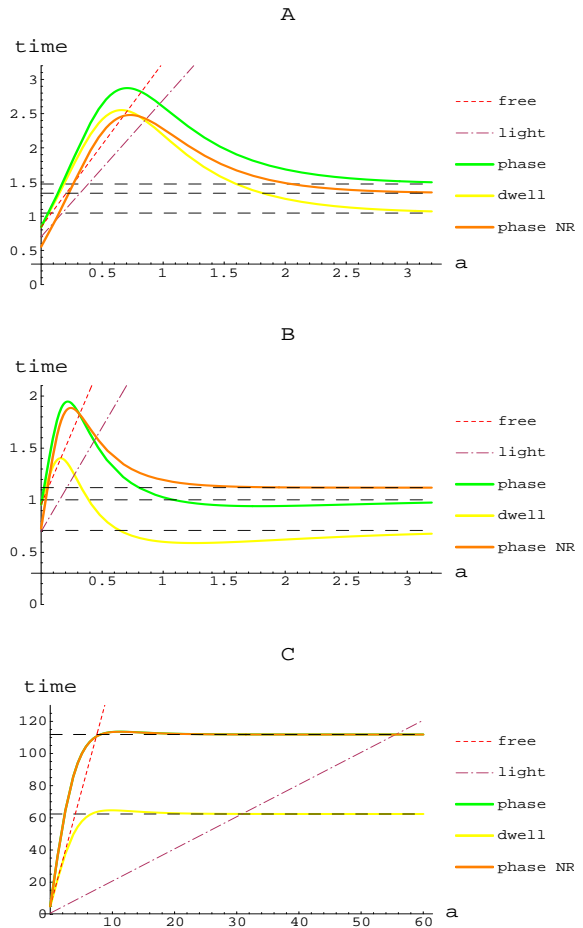}
\caption{(Color online) The phase and dwell times \emph{vs} barrier
width $a$. We used the system of \emph{natural units}, in which
$\hbar=c=1$. The energies were expressed in units of the particle
rest energy \emph{m} ($\backsimeq 0.5$ MeV for an electron, for
example). Accordingly, distances and times are given in units of
\emph{m}$^{-1}$. For reference we also show the \emph{free time} in
which the peak of a free wavepacket traverses the region $0<z<2a+l$
and the \emph{light time} in which a free light pulse traverses the
same region. For comparison we plotted also the nonrelativistic (NR)
phase time obtained from the Schr\"odinger equation \cite{ORS02}.
The horizontal dashed lines indicate the corresponding saturated
times in the opaque limit. The first two plots correspond to
energies in the relativistic scale, while the last one concerns to
nonrelativistic energies. In all these plots we take $l=0.7$. (A)
$E_0=1.8$, $V_0=1.5$; (B) $E_0=1.46$, $V_0=2.19$; (C) $E_0=1.01$,
$V_0=0.018$ (in this last plot the relativistic and the
nonrelativistic phase times coincide).\label{figure2}}
\end{figure}

\begin{figure}
\includegraphics[width=10cm,height=10cm]{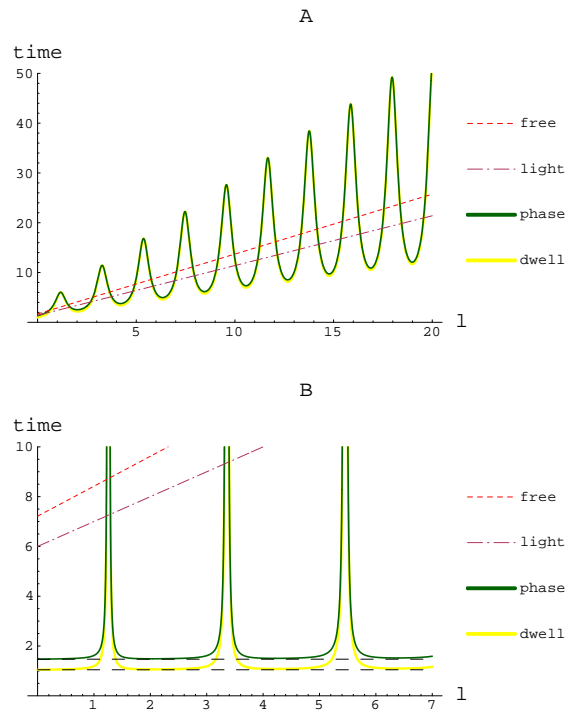}
\caption{(Color online) The phase and dwell times \emph{vs}
separation $l$ between the barriers. $E_0=1.8$ and $V_0=1.5$. (A)
$a=0.7$. (B) $a=3.0$. The horizontal dashed lines indicate the
corresponding saturated times in the opaque limit. We used the same
(natural) system of units as in the Figure 2. The peaks in both
figures correspond to resonant tunneling. \label{figure3}}
\end{figure}

In Figure \ref{figure3} we plot the behavior of $\tau_p$ and
$\tau_d$ for fixed $a$ and varying $l$. Figure \ref{figure3}A shows,
as expected, a trend to linear increase with $l$ (except for several
resonance peaks) as long as $a$ is not very large. That is, outside
the opaque domain both the dwell and phase times \textit{do not
saturate} with the barrier separation $l$. The same behavior is
observed for the nonrelativistic time with the same values of the
parameters (not shown in the figure). We can also observe the
equality between the dwell and phase times at resonance ($R\!=\!0$),
as predicted by the relation (\ref{dwell}). Again we observe that
the phase time off-resonance can be smaller than the light time,
even before attaining the saturated regime, which imply superluminal
group velocities. As the barrier width increases, the off-resonant
phase and dwell times tend to saturate to their values at the opaque
limit, but still presenting the resonant peaks, as we can observe in
Figure \ref{figure3}B. Finally, it is only when $a\to\infty$ that
both these times saturate in such a way that the resonant peaks are
no longer observed -- the generalized Hartmann effect. Therefore,
the results of the relativistic theory reinforce the conclusion by
Winful \cite{Win06-1} that the generalized Hartmann effect is just
an artifact resulting from taking the opaque limit before exploring
the variation with $l$ (for an alternative argument, using the
multiple peaks approach, see \cite{DLR05}).

Let us now look more carefully into the situation characterizing the
generalized Hartman effect by evaluating the flux of particles $J^z
= c\psi^\dag\alpha^z\psi$ both in the region between the barriers
(as given by $\psi_{III}$), and to the right of the potential for
the given energy $E_0$. The solution of the stationary problem gives
us
\begin{eqnarray}
C&=&\left[ \cosh(qa) + i\frac{(1-\alpha^2)}{2\alpha}\sinh
(qa)\right]
e^{ika}T \nonumber \\ \nonumber \\
D&=& - i\frac{(1+\alpha^2)}{2\alpha}\sinh (qa) e^{ik(3a+2l)}T\, ,
\nonumber
\end{eqnarray}
for the coefficients in $\psi_{III}$, with $T$ being the
transmission coefficient. From (\ref{t}) it is plain that in the
opaque limit $T\sim e^{-2qa}$, so that both $C$ and $D$ decay with
barrier width as $e^{-qa}$. Therefore, we conclude that in the
opaque limit, $qa\gg 1$, there is essentially no flux, hence, no
propagation of particles in regions $III$ and $V$. Accordingly, it
follows that the saturated times would be the same even if the
second barrier were absent \cite{WNL04}, similarly to what happens
in the nonrelativistic theory \cite{Win05}; in fact, expression
(\ref{p-op}) is identical to that obtained for the relativistic case
of a single barrier at the opaque limit (see \cite{KSG01} and
\cite{PJa03}). Thus, the condition to the generalized Hartman effect
to occur is the condition of no transmission, in which case it makes
no sense to associate any velocity to the tunneling process \cite{Win06-1}.

On the other hand, it is possible to have situations in which $qa$
is large, but finite, such that there is still an appreciable
transmission (before the saturation regime), and the associated
group velocities during the tunneling are superluminal. Notice that
we have considered wavepackets sharply centered around a given
energy in such a way that the transmitted wave packet could be seen
essentially as a (attenuated) nondistorted version of the incident
one, a feature that is claimed by several authors as allowing one to
attribute a physical meaning to the group velocity
\cite{distortionless} \cite{Win06-1}. Therefore, \textit{if} one
maintains the interpretation that the group velocities are
propagation velocities, it would seem that relativity does not
forbid superluminal tunneling velocities in the single or double
barrier tunneling. However, it must be noticed that while there is
little doubt that the group velocity in the region V (or region I,
for that matter) has a physical meaning, superluminal group
velocities emerge when we \emph{extrapolate} the concept of group
velocity to the region \emph{within} the barriers. But, it is clear
that inside the barriers the (evanescent) wavepacket undergoes great
distortion, not sharing the same shape as the incident or
transmitted ones; in fact, within the barriers the wavepacket does
not even have a peak which travels from one boundary to the another
\cite{KSG01, Win05, Win06-1}. Thus, despite the fact that the group
velocity has well defined meaning for the incident (before
reflections) and transmitted regions, the extrapolation of this
concept to the region inside the barriers cannot be justified, and
consequently there is no justification for associating this
(extrapolated) group velocity with tunneling velocity.

In what concerns the dwell time, since it does does not distinguish
between the transmitted and reflected channels, it is better
interpreted as a cavity lifetime. However, it is important to notice
that here, contrary to what happens for a Fabry-Perot cavity
\cite{Win05, Win06-1}, the phase and dwell times are not equal
in the off-resonance case [see Figs. 2 and 3 and the limits
(\ref{p-op})-(\ref{d-op})]. This prevents an immediate
identification of phase time as a cavity lifetime in the present
scenario.

Summarizing, in this paper we have analyzed the relativistic tunneling of a spin
1/2 particle through two successive electrostatic potential barriers
and showed that the so called generalized Hartmann effect also
occurs in the realm of relativistic quantum mechanics. In addition,
we obtained that the dwell time also saturates with the width of the
barriers. We demonstrated that the phase and dwell times can become
smaller than the light time (which implies superluminal group
velocities) even before the saturated regime is obtained, and we
observed that the group velocities predicted by the relativistic
theory can be smaller or greater than those predicted by
Schr\"odinger's theory, depending on the values of the parameters.
We also showed that the phase and dwell times show an almost linear
increasing with the separation between the barriers, and tends to
saturate \textit{only} when the barrier becomes extremely opaque.
Finally, we discussed a possible interpretation of the results,
favoring the argument that the group velocity cannot be interpreted
as a tunneling velocity.

\begin{acknowledgments}
JTL thanks CNPq/Brazil for partial support (Grant 201183/2005-6 PDE)
and also the Observatory of Complex Systems of the University of
Palermo for the kind hospitality.
\end{acknowledgments}

\appendix*
\section{}

In this Appendix we list the explicit expressions for the phase time, $\tau_p$, and the
self-interference delay, $\tau_i$. From (\ref{tp}) and (\ref{fase}) we obtain
\begin{equation}
\tau_p = \frac{1}{\hbar c^2}\left\{ (kl)\frac{E}{k^2} - \frac{1}{k^2q^2}\frac{h_1}{\Gamma^2 + \Delta^2}\right\}\; ,
\end{equation}
where we have defined
\begin{widetext}
\begin{eqnarray}
&&\Gamma\equiv 8 \alpha^2 \cosh (2qa) - 4(1+\alpha^2)^2 \sin^2 (kl) \sinh^2 (qa)\; , \\ \nonumber \\
&&\Delta\equiv 4 \alpha(1 - \alpha^2)\sinh (2qa) + 2(1+\alpha^2)^2\sin (2kl) \sinh^2 (qa) \; ,
\end{eqnarray}
and
\begin{eqnarray}
h_1 &\equiv& \Delta\left\{ 2(1+\alpha^2)\left[ (1+\alpha^2)Eq^2 (2kl)\sin (2kl) - 4\alpha^2 m c^2 (k^2+q^2)\cos (2kl)\right] \sinh^2 (qa) \right. \nonumber \\ \nonumber \\
&-& \left. 4\alpha^2 mc^2 (k^2+q^2)\left[ (1+\alpha^2) + (3-\alpha^2)\cosh (2qa) \right] \right. \nonumber \\ \nonumber \\
&+& \left. k^2 (2qa) (E-V_0) \left[ (1+\alpha^2)^2 \cos (2kl) - (1-6\alpha^2 +\alpha^4)\right]\sinh (2qa) \right\} \nonumber \\ \nonumber \\
&+& \Gamma \left\{-4\alpha(1-\alpha^2)k^2(2qa)(E-V_0)\cosh (2qa) \right. \nonumber \\ \nonumber \\
&+& \left. 2(1+\alpha^2)\left[ (1+\alpha^2)Eq^2(2kl) \cos (2kl) + 4\alpha^2 mc^2(k^2+q^2)\sin (2kl)\right]\sinh^2 (qa) \right. \nonumber \\ \nonumber \\
&+& \left. \left[ 4\alpha (1-3\alpha^2)mc^2 (k^2+q^2) -(1+\alpha^2)^2k^2 (2qa)(E-V_0)\sin (2kl)\right]\sinh (2qa) \right\} \; .
\end{eqnarray}

From equations (\ref{t}) and (\ref{r}), the self-interference delay,
as defined in (\ref{si}), is given by
\begin{equation}
\tau_i = \frac{m}{\hbar k^2}\frac{(1+\alpha^2)}{4 \alpha^3}\frac{h_2}{h_3}\; ,
\end{equation}
with
\begin{eqnarray}
h_2 &\equiv& \frac{1}{2} \alpha (1-\alpha^2)\sin (2kl) \sinh^2 (2qa) + \alpha^2 \cos^2 (kl) \sinh (4qa) \nonumber \\ \nonumber \\
&+& \alpha(1-\alpha^2) \sin (2kl) \sinh^2 (qa) \cosh (2qa) + (1-\alpha^2)^2 \sin^2 (kl) \sinh^2 (qa) \sinh (2qa) \; ,\\ \nonumber \\
h_3 &\equiv& \frac{1}{8\alpha^4} \left\{ 8 \alpha^4 \cosh^4 (qa) + \left[ 1 + 6\alpha^4 + \alpha^8 - (1-\alpha^4)^2 \cos (2kl) \right] \sinh^4 (qa) \right. \nonumber \\ \nonumber \\
&+& \left. \alpha^2 \left[ (1-\alpha^2)^2 + (1+\alpha^2)^2\cos (2kl) \right]\sinh^2 (2qa) + 2\alpha (1-\alpha^2)(1+\alpha^2)^2\sin (2kl) \sinh^2 (qa) \sinh (2qa) \right\} \; .
\end{eqnarray}
Finally, the dwell time is obtained from the phase time and the self-interference delay from (\ref{dwell}).

\end{widetext}


%

\end{document}